\documentclass[pdflatex,sn-mathphys-ay]{sn-jnl}

\usepackage[T1]{fontenc}          
\usepackage{xcolor}              
\usepackage{graphicx}            
\usepackage{adjustbox}           
\usepackage{booktabs}            
\usepackage{multirow}            
\usepackage{textcomp}            
\usepackage{amsmath,amssymb,amsfonts} 
\usepackage{bm}                  
\usepackage{mathrsfs}            
\usepackage{amsthm}              
\usepackage{etoolbox}            
\usepackage{appendix}            
\usepackage{algorithm}           
\usepackage{algorithmicx}        
\usepackage{algpseudocode}       
\usepackage{listings}            
\usepackage{subfig}             
\usepackage{manyfoot}            
\usepackage{threeparttable}

\DeclareMathOperator{\Prob}{\mathbb{P}}

\theoremstyle{thmstyleone}%
\theoremstyle{thmstyletwo}%

\theoremstyle{thmstylethree}%

\begin{document}

\title[Finite Mixture Cox Model for Heterogeneous Time-dependent Right-Censored Data]{Finite Mixture Cox Model for Heterogeneous Time-dependent Right-Censored Data}


\author*[1]{\fnm{Ahmad} \sur{Talafha}}\email{atalafha@stedwards.edu}



\affil*[1]{\orgdiv{Department of Mathematics}, \orgname{St. Edward's University}, \orgaddress{\street{3001 S Congress Ave}, \city{Austin}, \postcode{78704}, \state{TX}, \country{USA}}}




\abstract{In this study, we address the challenge of survival analysis within heterogeneous patient populations, where traditional reliance on a single regression model such as the Cox proportional hazards (Cox PH) model often falls short. Recognizing that such populations frequently exhibit varying covariate effects, resulting in distinct subgroups, we argue for the necessity of using separate regression models for each subgroup to avoid the biases and inaccuracies inherent in a uniform model. To address subgroup identification and component selection in survival analysis, we propose a novel approach that integrates the Cox PH model with dynamic penalty functions, specifically the smoothly clipped absolute deviation (SCAD) and the minimax concave penalty (MCP). These modifications provide a more flexible and theoretically sound method for determining the optimal number of mixture components, which is crucial for accurately modeling heterogeneous datasets. Through a modified expectation--maximization (EM) algorithm for parameter estimation and component selection, supported by simulation studies and two real data analyses, our method demonstrates improved precision in risk prediction.}



\maketitle

\section{Introduction}\label{sec1}
While survival analysis has made notable strides and offers significant advantages in various fields, it's often incorrectly assumed that the conditional hazard function for a heterogeneous population, given certain covariates, can be accurately represented by a single regression model, such as the Cox proportional hazards (Cox PH) model. This assumption, however, may not hold true in practice, especially in cases of heterogeneity, evidenced by the varying effects of covariates, leading to the natural formation of different subgroups within the population. In these instances, it is more appropriate to use separate regression models for each subgroup, rather than a single uniform model. Relying on classical Cox regression analysis in these circumstances can result in biased estimates and incorrect conclusions. For example, \cite{yan_2021} utilized a concave fusion penalty for subgroup identification and parameter estimation with censored data. However, this method does not offer a posterior probability for subgroup membership, thus limiting its predictive capabilities. As an alternative, structured mixture models have been used to mitigate this limitation, exemplified by the structured logistic normal mixture model proposed by \cite{shen_he_2015} and further extended by \cite{wu_2016} to a logistic-Cox mixture model accommodating censored outcomes. \cite{you_2018} explored variable selection in finite-mixture Cox models, while \cite{Nagpal_2021} tackled some numerical challenges in fitting these models, suggesting a default of three latent subgroups. Yet, these mixture model approaches require predetermined specification of the number of mixing components, which may not always be practical. Component selection is an important issue in finite mixture models. Determining the correct number of components is essential for estimation, inference, and interpretation. Current component selection techniques include methods based on the likelihood ratio test (\cite{chen_1996_Penalized,Kasahara_2015_testing,Li_2010_testing,lo_2005_likelihood,lo_2001_testing}), distance measure based methods ( \cite{james_2001_consistent,Woo_2006_robust}), and penalty based approaches (\cite{chen_1996_Penalized,Chen_2008_order,Huang_2017}). However, these methods were developed for parametric mixture models, i.e., mixtures of parametric density models or mixtures of linear regression models, and may not be applicable to models with semi-parametric mixture components. Also, the information criteria, i.e., the Akaike information criterion
(AIC) (\cite{Akaike1998}) and the Bayesian information criterion (BIC) (\cite{Schwarz_1978_Estimating}), provide a general component-selection framework for mixture models consisting of parametric or nonparametric models.

Addressing a similar concern in the realm of credit risk, \cite{Pei_2022} proposed a latent class Cox model for heterogeneous time-to-event data with penalty that was proposed initially by \cite{Huang_2017}. The principal contribution of our approach is in the modification of the penalty term within the framework of the extended Cox proportional hazards model. \cite{Pei_2022} contribution lies in extending the Cox model to more adeptly handle heterogeneous time-to-event data, specifically by accommodating covariate effect heterogeneity without necessitating predefined latent classes. Our work builds on this by proposing the model's dynamic penalty function for component selection.

Specifically, we replace the mixing proportion \(\pi\) with a dynamic penalty function \(p_{\lambda}(\pi)\), where \(p_{\lambda}\) denotes the smoothly clipped absolute deviation (SCAD) (\cite{Fan_2001}) or the minimax concave penalty (MCP) (\cite{Zhang_2010}). This modification introduces a more flexible and theoretically grounded mechanism for determining the optimal number of mixture components within the model. The SCAD and MCP penalties are chosen for their established advantages in variable selection methods, known for their desirable theoretical properties. By integrating \(p_{\lambda}(\pi)\) into the penalty function, our model improves the precision of component selection, essential for accurately modeling the underlying cox  PH model in heterogeneous datasets.

The rest of the paper is organized as follows. In Section \ref{sec:meth}, we first introduce the finite mixture Cox
model based on SCAD penalty and MCP. A penalized likelihood approach for the identification and estimation of the model is proposed, and a modified
EM algorithm is used to implement it. A BIC-type criterion is also proposed to select the tuning parameters of the SCAD and MCP. Section \ref{sec:model_eval} includes model evaluation. The numerical results, including a simulation study and two real
data application, are presented in Sections \ref{sec:numerical_study} and \ref{sec:real_data}. Section \ref{sec:conclusion} provides some concluding remarks and discusses possible future
work. Derivation of the proposed method is relegated to Appendix \ref{appendixA}.

\section{The latent class Cox PH model}
\label{sec:meth}

In the context of time-to-event analysis, we are often interested in studying the time until an event of interest occurs, which could range from the onset of a disease, mortality, breast cancer diagnosis, HIV (human immunodeficiency virus) infection, to the occurrence of a particular life event. Censoring, where the event of interest has not occurred at the end of the study period, presents a significant analytical challenge. Let $T_i$ denote the time until the event occurs for subject $i$, and $C_i$ represent the censoring time. The survival function, $S_i(t) = \Prob(T_i > t)$, then quantifies the probability that the event has not occurred by time $t$. Observations consist of the triplet $(\bm{X}_i, Y_i, \delta_i)$, where $Y_i = \min(T_i, C_i)$ is the observed time until the event or censoring, and $\bm{X}_i = (X_{i1}, \ldots, X_{ip})^\top$ includes covariates of the subject. The indicator $\delta_i = I(T_i \leq C_i)$ distinguishes whether the event of interest was observed $(\delta=1)$ within the study period or censored $(\delta=0)$.

The Cox proportional hazards model provides a framework for relating the hazard of the event, $\lambda_i(t) = \lambda_0(t) \exp(\bm{\beta}^\top \bm{X}_i)$, to subject-specific covariates $\bm{X}_i$, where $\lambda_0(t)$ is the baseline hazard function, and $\bm{\beta} = (\beta_1, \ldots, \beta_p)^\top$ measures the effect of covariates on the event's hazard. This model assumes a homogeneous effect of covariates across the study population. Note that $(\cdot)^\top$ signifies the transpose of a vector or matrix. Recognizing potential heterogeneity, a group-specific Cox model can be employed, $\lambda_k(t) = \lambda_{0k}(t) \exp(\bm{\beta}_k^\top \bm{X}_i)$, to accommodate different subpopulations, with $\bm{\beta}_k = (\beta_{k,1}, \ldots, \beta_{k,p})^\top$ representing group-specific covariate effects.

Considering the possibility that data originates from a mixture of different subpopulations, the model can be extended to a mixture model with proportions $\bm{\pi}=(\pi_1, \ldots, \pi_K)$, where $\pi_k$ denotes the prior probability of a subject belonging to the $k^{th}$ component. The model's conditional density is formulated as $f(y, \delta|\bm{x}) = \sum\limits_{k=1}^{K} \pi_k f_k(y, \delta|\bm{x})$, with each component having a specific conditional density. For the Cox PH model, the conditional density function for group $k$ is
\begin{align} \label{eq:group_k_conditional_density}
f_k(y, \delta|\mathbf{x}) = \left[ \lambda_{0k}(y) \exp \left( \bm{\beta}_{k}^{\top} \mathbf{x} \right) \right]^{\delta} \exp \left\{ -\exp \left( \bm{\beta}_{k}^{\top} \mathbf{x} \right) \int_0^y \lambda_{0k}(u)du \right\}.
\end{align}

For identifiability, it is necessary that $0 < \pi_k < 1$ for all $k$ and $\sum\limits_{k=1}^{K} \pi_k = 1$, and that the parameter vectors $\bm{\beta}_1, \bm{\beta}_2, \ldots, \bm{\beta}_K$ are distinct implying at least one differing component between $\bm{\beta}_k$ and $\bm{\beta}_{k'}$ for $k \neq k'$.

The log-likelihood based on the observed data, $\ell_n(\theta; Y, \Delta | X) = \sum\limits_{i=1}^{n} \log \left( \sum\limits_{k=1}^{K} \pi_k f_k(Y_i, \delta_i | X_i) \right)$, underpins the parameter estimation process. This process may leverage the expectation–-maximization (EM) algorithm to accommodate latent subgroup memberships, thereby facilitating the estimation of both the mixture proportions $\bm{\pi}$ and the covariate effects $\bm{\beta}$. Here, $\bm{Y} = (Y_1, \ldots, Y_n)^\top$, $\Delta = (\delta_1, \ldots, \delta_n)^\top$, and $\bm{X} = (X_1, \ldots, X_n)^\top$ aggregate the times to event or censoring, censoring indicators, and covariate information for all subjects, respectively.

Let $s_{ik}$ denote the latent indictor variable according to whether or not the $i$-th sample comes from the $k$-th subpopulation, and let $\bm{S}=\left(s_{ik} \right)_{n \times K}$. The complete log-likelihood 
\begin{align}
    \ell_{c}\left(\bm{\theta}; \bm{Y}, \Delta, \bm{S} | \bm{X} \right) = \sum\limits_{i=1}^{n} \sum\limits_{k=1}^{K} s_{ik} \log\left(\pi_{k} \right) + \sum\limits_{i=1}^{n} \sum\limits_{k=1}^{K} s_{ik} \log\left(f_{k}(Y_{i}, \delta_{i} | \bm{X}_{i} )\right)
    \label{eq:complete_log_likelihood}
\end{align}

The maximum likelihood estimates of $\bm{\pi}$ and $\bm{\beta}$ are denoted $\widehat{\bm{\pi}}$ and $\widehat{\bm{\beta}}$, respectively. Given $\widehat{\bm{\pi}}$ and $\widehat{\bm{\beta}}$, the conditional probabilities of $s_{ik}$ can be solved by the EM algorithm.
\subsection{Penalized likelihood estimation}

The expected complete-data log-likelihood, as detailed in equation (\ref{eq:complete_log_likelihood}), significantly involves $\log(\pi_{k})$. The gradient of this term increases rapidly as $\pi_{k}$ approaches zero, presenting challenges for traditional $L_{p}$ penalties in setting insignificant $\pi_{k}$ values to zero. Analysis by \cite{Huang_2017} reveals that $L_{1}$ penalties do not globally address this issue. To enhance sparsity in $\bm{\pi}$, it is proposed to apply penalties directly to $\log(\pi_{k})$.

\cite{Huang_2017} recommends penalizing $\log(\varepsilon + \pi_k) - \log(\varepsilon)$ in the Gaussian mixture model, where $\varepsilon$ is a small positive constant. This approach, also adopted for the Cox proportional hazard model by \cite{Pei_2022}, ensures that the penalty function is monotonically increasing and becomes negligible for small $\pi$ values; for ease of reference and comparison in our analysis, we term this penalization approach as the logarithmic scale (LS) penalty. The penalized likelihood function is expressed as:
\begin{equation}
\ell_P (\rho,\bm{\theta}) = \sum_{i=1}^{n} \sum_{k=1}^{K} s_{ik} \left[ \log(\pi_k) + \log f_k (Y_i, \delta_i|\mathbf{X}_i) \right] - n \rho \sum_{k=1}^{K} \left[\log (\varepsilon + \pi_k) - \log(\varepsilon)\right].
\end{equation}

According to \cite{Fan_2001}, an ideal penalty function should exhibit unbiasedness, sparsity, and continuity. To address the potential bias induced by the LS penalty for large $\pi_{k}$ values, we explore the smoothly clipped absolute deviation (SCAD) function. The SCAD penalized likelihood function is given by:
\begin{equation}
\ell_P (\kappa,\bm{\theta}) = \sum_{i=1}^{n} \sum_{k=1}^{K} s_{ik} \left[ \log(\pi_k) + \log f_k (Y_i, \delta_i|\mathbf{X}_i) \right] - n \kappa \sum_{k=1}^{K} \left[\log (\varepsilon + p^{\text{SCAD}}_{\kappa,a}(\pi_k)) - \log(\varepsilon)\right],
\label{eq:SCAD_log_likelihood}
\end{equation}
where the SCAD penalty and its derivative, as proposed by \cite{Fan_2001}, are defined as:
\begin{align}
    p'^{\, \text{SCAD}}_{\kappa,a}(\pi) = I(\pi \leq \kappa) + \frac{(a\kappa - \pi)_+}{(a - 1)\kappa} I(\pi > \kappa),
\end{align}
with $a > 2$ and $(x)_+ = \max\{0, x\}$.

The minimax concave penalty (MCP) penalized likelihood function is given by:
\begin{equation}
\ell_P (\alpha,\bm{\theta}) = \sum_{i=1}^{n} \sum_{k=1}^{K} s_{ik} \left[ \log(\pi_k) + \log f_k (Y_i, \delta_i|\mathbf{X}_i) \right] - n \alpha \sum_{k=1}^{K} \left[\log (\varepsilon + p^{\text{MCP}}_{\alpha,b}(\pi_k)) - \log(\varepsilon)\right],
\label{eq:MCP_log_likelihood}
\end{equation}

where MCP is described by its derivative as follows (\cite{Zhang_2010}):
\begin{align}
    p'^{\, \text{MCP}}_{\alpha,b}(\pi) = \left( \alpha-\frac{\pi}{b}\right)I(\pi \leq b \alpha),
\end{align}
where $b > 1$. This function linearly increases up to a threshold and then remains constant, effectively penalizing large coefficients without inducing bias.

Both the SCAD and MCP penalties adhere to the criteria set by \cite{Fan_2001}, striking a balance between unbiasedness, sparsity, and continuity. In \cite{Fan_2001} work, the SCAD penalty is approached through a local quadratic approximation, a method that can lead to instability when the estimated values are near zero. Drawing from \cite{Hunter_2005_variable}, our method applies the concept of perturbation to the design of the penalty function. Furthermore, we employ a local linear approximation (\cite{zou_2008}), to both achieve effective shrinkage of the mixture probability and ensure the stability of the estimation process. A similar approach is adopted for implementing the MCP penalty in our method. 

\subsection{Likelihood Estimation via Expectation-Maximization Algorithm}

\subsubsection{Expectation Step}
This step calculates the expectation of the penalized complete-data log-likelihood, equations (\ref{eq:SCAD_log_likelihood}) and (\ref{eq:MCP_log_likelihood}), considering the hidden class labels $\bm{S}$, given the observed data $D = (\mathbf{X}, Y, \Delta)$ and current parameter estimates $\bm{\theta}^{(m)}$, where $m$ denotes the iteration number. The expectation for SCAD and MCP penalized likelihoods are denoted as $Q_{SCAD}$ and $Q_{MCP}$, respectively:
\begin{align*}
& Q_{SCAD}(\kappa, a, \bm{\theta}; \bm{\theta}^{(m)}) = \mathbb{E} \left[ \ell_P ( \kappa,\bm{\theta}) \mid D; \bm{\theta}^{(m)} \right] \\
&= \sum_{i=1}^{n} \sum_{k=1}^{K} \mathbb{E} \left[ s_{ik} \mid D; \bm{\theta}^{(m)} \right] \log \left[ \pi_k f_k (Y_i, \delta_i \mid \mathbf{X}_i) \right]- n \,  \kappa \sum_{k=1}^{K} \left[\log (\varepsilon + p^{\text{SCAD}}_{\kappa,a}(\pi_k)) - \log(\varepsilon)\right], \\
& Q_{MCP}(\alpha, b, \bm{\theta}; \bm{\theta}^{(m)}) = \mathbb{E} \left[ \ell_P (\alpha, \bm{\theta}) \mid D; \bm{\theta}^{(m)} \right] \\
&= \sum_{i=1}^{n} \sum_{k=1}^{K} \mathbb{E} \left[ s_{ik} \mid D; \bm{\theta}^{(m)} \right] \log \left[ \pi_k f_k (Y_i, \delta_i \mid \mathbf{X}_i) \right] - n \, \alpha \sum_{k=1}^{K} \left[\log (\varepsilon + p^{\text{MCP}}_{\alpha,b}(\pi_k)) - \log(\varepsilon)\right],
\end{align*}
where $s^{(m)}_{ik}$ is the posterior probability that the data are generated from cluster $k$, calculated as:
\begin{align}
s^{(m)}_{ik} = \frac{\pi^{(m)}_k f_k \left( Y_i, \delta_i \mid \mathbf{X}_i \right)}{\sum\limits_{k=1}^{K} \pi^{(m)}_k f_k \left( Y_i, \delta_i \mid \mathbf{X}_i \right)}.
\label{eq:posterior_probability}
\end{align}

\subsubsection{Maximization Step}
This step seeks to update the parameter vector $\bm{\theta}$ by maximizing the $Q$ functions, $Q_{SCAD}$ and $Q_{MCP}$, thus updating the parameter vector to $\bm{\theta}^{(m+1)} = \arg \max_{\bm{\theta}} Q(\bm{\theta}; \bm{\theta}^{(m)})$. The decompositions of the $Q$ functions for SCAD and MCP are as follows:
\begin{align}
Q_{SCAD}(\kappa, a, \bm{\theta}; \bm{\theta}^{(m)}) &= Q_{SCAD}(\kappa,a, \bm{\pi}; \bm{\theta}^{(m)}) + \sum_{k=1}^{K} Q_{2,k}(\bm{\beta}_{k}; \bm{\theta}^{(m)}), \\
Q_{MCP}(\alpha, b, \bm{\theta}; \bm{\theta}^{(m)}) &= Q_{MCP}(\alpha, b, \bm{\pi}; \bm{\theta}^{(m)}) + \sum_{k=1}^{K} Q_{2,k}(\bm{\beta}_{k}; \bm{\theta}^{(m)}),
\end{align}
where $Q_{SCAD}$ and $Q_{MCP}$ for mixing proportions $\bm{\pi}$ are updated by maximizing:
\begin{align}
Q_{SCAD}(\kappa, a, \bm{\pi}; \bm{\theta}^{(m)}) &= \sum_{i=1}^{n} \sum_{k=1}^{K} s^{(m)}_{ik} \log(\pi_k) - n \, \kappa \sum_{k=1}^{K} \left[\log (\varepsilon + p^{\text{SCAD}}_{\kappa,a}(\pi_k)) - \log(\varepsilon)\right], \\
Q_{MCP}(\alpha, b, \bm{\pi}; \bm{\theta}^{(m)}) &= \sum_{i=1}^{n} \sum_{k=1}^{K} s^{(m)}_{ik} \log(\pi_k) - n \, \alpha \sum_{k=1}^{K} \left[\log (\varepsilon + p^{\text{MCP}}_{\alpha,b}(\pi_k)) - \log(\varepsilon)\right].
\label{eq:Q_SCAD_MCP}
\end{align}

and
\begin{align}
Q_{2,k}(\bm{\beta}_{k}; \bm{\theta}^{(m)}) &= \sum_{i=1}^{n} s^{(m)}_{ik} \log \left[ f_k(Y_i, \delta_i \mid \mathbf{X}_i) \right],
\label{eq:Q2k_function}
\end{align}

The parameter updates for $\bm{\beta}_k$ are derived by maximizing $Q_{2,k}(\bm{\beta}_{k}; \bm{\theta}^{(m)})$. For numerical stability and to ensure sparsity in the mixing proportions, any proportions smaller than a predefined threshold $\tilde{\varepsilon}$ are set to zero.

It follows that the maximization of the $Q$ functions can be done by maximizing separately $$Q_{SCAD}(\kappa,a, \bm{\pi}; \bm{\theta}^{(m)})$$ and $$ Q_{MCP}(\alpha,b, \bm{\pi}; \bm{\theta}^{(m)})$$ with respect to the mixing proportions, and for each component $k$, maximizing $Q_{2,k}(\bm{\beta}_k; \bm{\theta}^{(m)})$ with respect to the regression parameters $\bm{\beta}_k$.
The mixing proportions are updated by maximizing either of (\ref{eq:Q_SCAD_MCP}) with respect to the mixing proportions $\bm{\pi}$ subject to the constraint $\sum\limits_{k=1}^{K} \pi_k = 1$. Hence, we introduce Lagrange multiplier $\rho_{1}$ and $\rho_{2}$ to take into account the constraint and aim to solve the following set of equations:
\begin{align}
\frac{\partial}{\partial \pi_k} \left[ \sum_{i=1}^{n} \sum_{k=1}^{K} s^{(m)}_{ik} \log(\pi_k) - n \, \kappa \sum_{k=1}^{K} [\log (\varepsilon + p^{\text{SCAD}}_{\kappa,a}(\pi_k))- \rho_{1} \left( \sum_{k=1}^{K} \pi_k - 1 \right) \right] = 0.
\label{eq:mixing_proportions_update_SCAD}
\end{align}
gives
\begin{align}
\hat{\pi}^{\text{SCAD} \, (m+1)}_k = \frac{\sum\limits_{i=1}^{n} s^{(m+1)}_{ki}}{ n-n \kappa \sum\limits_{k=1}^{K} \frac{p'^{\, \text{SCAD}}_{\kappa,a}(\pi^{(m)}_{k}) \, \pi^{(m)}_{k}}{\varepsilon+p^{\text{SCAD}}_{\kappa,a}(\pi^{(m)}_{k}) }  + n \kappa  \frac{p'^{\, \text{SCAD}}_{\kappa,a}(\pi^{(m)}_{k})}{\varepsilon+p^{\text{SCAD}}_{\kappa,a}(\pi^{(m)}_{k}) }}
\label{eq:pi_k_scad}
\end{align}

Similarly, for MCP penalty, 

\begin{align}
\hat{\pi}^{\text{MCP} \, (m+1)}_k = \frac{\sum\limits_{i=1}^{n} s^{(m+1)}_{ki}}{ n-n \alpha \sum\limits_{k=1}^{K} \frac{p'^{\, \text{MCP}}_{\alpha,b}(\pi^{(m)}_{k}) \, \pi^{(m)}_{k}}{\varepsilon+p^{\text{MCP}}_{\alpha,b}(\pi^{(m)}_{k}) }  + n \alpha  \frac{p'^{\, \text{MCP}}_{\alpha,b}(\pi^{(m)}_{k})}{\varepsilon+p^{\text{MCP}}_{\alpha,b}(\pi^{(m)}_{k}) }}
\label{eq:pi_k_mcp}
\end{align}

The $(m + 1)$st M step updates about $\bm{\beta}_k$ by maximizing (\ref{eq:Q2k_function}) for $k = 1, 2, \ldots, K$, separately. The mixing proportion is hard to shrink to exactly zero in numerical studies. Similar to \cite{Huang_2017}, we use a reasonably small pre-specified threshold $\tilde{\varepsilon}$, i.e., $10^{-5}$, and set the mixing proportions smaller than the threshold to zero.

In order to maintain the flow of our discussion, the detailed derivation of $\hat{\pi}^{\text{SCAD} \, (m+1)}_k$, which is also similar to $\hat{\pi}^{\text{MCP} \, (m+1)}_k$ has been relegated to Appendix \ref{appendixA}. 

Note that the density function in (\ref{eq:Q2k_function}) contains a nonparametric function $\lambda_{0k}(\cdot)$, we adopt the profile likelihood approach (\cite{Johansen_1983_extension} ) to update it. Denote $\lambda_{0k}(Y_i)$ as $\lambda_{i,k}$ at time $Y_i$. Maximizing over $\lambda_{i,k}$, we can obtain profile estimates of the hazards as a function of $\bm{\beta}_k$. 
\begin{align}
\hat{\lambda}_{i,k} = \frac{s^{(m)}_{ik}}{\sum\limits_{j:Y_j \geq Y_i} s^{(m)}_{jk} \exp \left( \bm{\beta}^{\top}_k X_j \right)}. 
\label{eq:lambda_i_k_estimate}
\end{align}
Replacing $\lambda_{0k}(Y_i)$ in (\ref{eq:Q2k_function}) with $\hat{\lambda}_{i,k}$ yields the following weighted log-partial likelihood:
\begin{align}
\ell_{c,k} (\bm{\beta}_k;D, Z^{(m)}) = \sum_{i=1}^{n} \delta_i s^{(m)}_{ik} \left[ \bm{\beta}^{\top}_k X_i - \log \left( \sum_{j:Y_j \geq Y_i} s^{(m)}_{jk} \exp \left( \bm{\beta}^{\top}_k X_j \right) \right) \right]. 
\label{eq:weighted_partial_likelihood}
\end{align}
Moreover, the estimates in (\ref{eq:lambda_i_k_estimate}) are discrete and cannot be used directly to estimate the density function in (\ref{eq:group_k_conditional_density}). Thus, following \cite{Bordes_2016_Stochastic}, we apply a kernel smoothing technique to obtain a smooth estimator for the baseline hazard function $\lambda_{0k}(\cdot)$. Suppose that $\mathcal{K}(\cdot)$ is a kernel function and $h = h_n$ is a bandwidth; then $\lambda^{(m+1)}_{0k}(t)$ is estimated by
\begin{align}
\hat{\lambda}^{(m+1)}_{0k}(t) = \frac{1}{h} \sum_{i=1}^{n} \mathcal{K} \left( \frac{t - Y_i}{h} \right) \hat{\lambda}_{i,k}.
\label{eq:smooth_baseline_hazard}
\end{align}
$\lambda_{0k}(\cdot)$ is then updated by (\ref{eq:smooth_baseline_hazard}). Algorithm \ref{Algorithm:EM} summarizes the proposed modified EM algorithm.  Details of (\ref{eq:lambda_i_k_estimate}) derivation is given in Appendix \ref{appendixA}
\begin{algorithm}[h!]
\caption{Modified EM Algorithm for Finite Cox PH Mixture Models}
\begin{algorithmic}
\State Initialize: set $m = 0$
\State Initialize $\bm{\pi}^{(0)}$ and $\bm{S}^{(0)}$
\State Estimate $\bm{\beta}^{(0)}$ by maximizing (\ref{eq:weighted_partial_likelihood})
\State Estimate $\lambda_{0k}^{(0)}(\cdot)$ according to (\ref{eq:smooth_baseline_hazard})
\Repeat
    \State \textbf{Step E:}
    \State Compute $\bm{S}^{(m+1)}$ according to (\ref{eq:posterior_probability})
    \State \textbf{Step M:}
    \State Update $\pi^{(m+1)}$ according to either (\ref{eq:pi_k_scad}) or (\ref{eq:pi_k_mcp})
    \State Delete component $k$ of which $\pi^{(m+1)}_{k} < \tilde{\varepsilon}$
    \For{$k = 1$ to $K$}
        \State Update $\bm{\beta}_k^{(m+1)}$ by solving (\ref{eq:weighted_partial_likelihood})
        \State Update $\lambda_{0k}^{(m+1)}$ according to (\ref{eq:smooth_baseline_hazard})
    \EndFor
    \State Set $m = m + 1$
\Until{convergence} 
\end{algorithmic}
\label{Algorithm:EM}
\end{algorithm}

\subsubsection{Selection of Tuning Parameters}

In our penalized component-selection framework, as outlined in equation (\ref{eq:SCAD_log_likelihood}), selecting the number of components under the SCAD penalty involves determining the tuning parameters $\kappa$ and $\alpha$. It has been observed that the performance of SCAD is relatively insensitive to variations in $a$ (\cite{Fan_2001}), a finding corroborated in our simulation studies for the model in question. Consequently, we adopt $a = 3.7$, following the recommendation by \cite{Fan_2001}. In this section, we detail our methodology for selecting the tuning parameter $\kappa$.

Drawing on established practices in component selection for mixture models \citep{Du_2013_Simultaneous, Huang_2017, wang_2007_Tuning, Pei_2022}, we introduce $\text{BIC}_{\text{SCAD}}(\kappa)$ as a criterion for selecting $\kappa$ in the context of the Cox mixture model:
\begin{align}
    \text{BIC}_{\text{SCAD}}(\kappa) = \sum\limits_{i=1}^{n} \log\left\{\sum\limits_{k=1}^{\hat{K}} \hat{\pi}_{k} \, f_{k}\left(Y_{i}, \delta_{i} | \bm{X}_{i}; \hat{\bm{\beta}}_{k} \right) \right\} -\frac{1}{2} C_{n} \, D_{f} \log(n),
\end{align}
where $D_{f} = \hat{K}-1+\hat{K}p$ represents the degrees of freedom for the proposed Cox proportional hazard model, with $\hat{K}$ as the estimate of the number of components, and $\hat{\pi}_{k}$, $\hat{\bm{\beta}}_{k}$ are the estimates obtained by maximizing equation (\ref{eq:Q_SCAD_MCP}). The term $C_n$ denotes a positive number that may vary with $n$. The modified BIC criterion converges to the conventional BIC (\cite{Schwarz_1978}) when $C_n = 1$. Following \cite{Ma_2017_concave}, who set $C_n = c \log(\log(n + p))$ in scenarios with increasing numbers of predictors $p$ alongside sample size $n$, we employ $C_n = c \log(\log(n + K))$, where $c$ is a positive constant. The optimal tuning parameter is then selected as:
\begin{align}
    \kappa_{\text{BIC}_{\text{SCAD}}} = \arg \max_\kappa \text{BIC}_{\text{SCAD}}(\kappa).
\end{align}

Similarly, for the MCP penalty, the optimal tuning parameter is selected as:
\begin{align}
    \alpha_{\text{BIC}_{\text{MCP}}} = \arg \max_\alpha \text{BIC}_{\text{MCP}}(\alpha),
\end{align}
with $b = 3$, as recommended by \cite{Zhang_2010}.





\subsubsection{Convergence Criteria}

The algorithm ends the iterations according to the following two criteria that hold simultaneously:
\begin{enumerate}
    \item Absolute convergence criterion: The modified EM algorithm is considered to have converged if the difference between the current log-likelihood function value $\ell(\bm{\theta}^{(t)})$ and the log-likelihood function value $\ell(\bm{\theta}^{(t-1)})$ of the previous iteration is less than some threshold $\epsilon$
    \begin{align}
        \left| \ell(\bm{\theta}^{(t)}) - \ell(\bm{\theta}^{(t-1)}) \right| < \epsilon. \label{eq:absolute_convergence} 
    \end{align}
    
    \item Relative convergence criterion: The modified EM algorithm is considered to have converged if the difference between the log-likelihood function value $\ell(\bm{\theta}^{(t)})$ and the log-likelihood function value $\ell(\bm{\theta}^{(t-1)})$ of the previous iteration divided by the log-likelihood function value is less than some threshold $\tilde{\epsilon}$
        \begin{align}
    \left| \frac{\ell(\bm{\theta}^{(t)}) - \ell(\bm{\theta}^{(t-1)})}{\ell(\bm{\theta}^{(t)})} \right| < \tilde{\epsilon}. 
    \label{eq:relative_convergence}
        \end{align}
\end{enumerate}

\subsubsection{Initial values}

One of the principal advantages of the EM algorithm is its guaranteed convergence property. Nonetheless, a significant drawback is that, contingent upon the initial values, the algorithm might converge to a local maximum instead of the desired global maximum. This issue is highlighted in the literature, where it is noted that the algorithm's convergence to a local optimum can depend heavily on the chosen starting points (\cite{Tanner_1993_Tools}). Furthermore, \cite{McLachlan_2000} have pointed out that selecting initial values near the boundary of the parameter space can lead to situations where convergence of the parameters is unattainable. Following \cite{Pei_2022}, initialize the values for the mixing probabilities, $\bm{\pi}^{(0)}$, and the component assignment matrix, $\bm{S}^{(0)}$, we start with a predefined large number of mixture components, specifically $K = 10$. Our proposed modified EM algorithm subsequently reduces the total number of components by iteratively merging smaller components into larger ones. The initialization of $\bm{S}^{(0)}$ involves creating an $n \times K$ matrix where, for each row, a single column in the range $1$ to $K$ is randomly selected to have its element set to $1$, with all other elements in the row set to $0$. This process leads to the initial mixing probability, $\bm{\pi}^{(0)}$, being determined by calculating the average of the elements across each row in $\bm{S}^{(0)}$, thereby setting each component of $\bm{\pi}^{(0)}$ to $\frac{1}{K}$.

\section{Model evaluation} \label{sec:model_eval}

In diagnostic screenings, markers are often employed to assess the likelihood of disease onset in individuals. These markers may be single continuous measures such as cell phase percentages for breast cancer detection (\cite{Heagerty2000}), CD4 cell counts for AIDS identification (\cite{Hung2010a}), or HIV-1 RNA levels to recognize HIV (\cite{Song2012}). Additionally, scores generated from regression models involving multiple risk factors can serve as markers. For instance, \cite{Chambless2006} used logistic regression scores including an array of traditional and new risk factors for coronary heart disease (CHD) prediction, while \cite{Lambert2014} utilized prognostic scores based on several covariates to forecast survival in patients with compensated cirrhosis and acute leukemia. Published scores, such as the Framingham risk score for cardiovascular patients (\cite{Cai2006}) and the Karnofsky score for lung cancer patients, are also used as markers \cite{Kalbfleisch2011}. In our case risk scores (markers) are obtained either as linear predictor from the fitted Cox model or  linear predictor from the fitted Cox mixture model.

Let $T_{i}$ denote the time of disease onset and $M_{i}$ represent the marker value for individual $i$, where $i = 1,\ldots,n$. Let $D_{i}(t)$ indicate the disease status at time $t$, taking values of $1$ or $0$. Given a threshold $c$, the time-dependent sensitivity and specificity  are respectively defined by
\[ \operatorname{Sensitivity}(c, t) = P(M_{i} > c \, | \,  T \leq c) = P(M_{i} > c | D_{i}(t) = 1) \]
\[ \operatorname{Specificity}(c, t) = P(M_{i}
\leq c \,  | \, T > c ) = P(M_{i}
\leq c \, | D_{i}(t) = 0)\]
From these definitions, the receiver operating characteristic (ROC) curve at any time $t$ can be denoted as $\operatorname{ROC}(t)$, which plots $\operatorname{Sensitivity}(c,t)$ against $1-\operatorname{Specificity}(c,t)$ for different thresholds $c$. The time-dependent 
area under the ROC curve (AUC) is expressed as
\[ AUC(t) = \int_{-\infty}^{\infty} \operatorname{Sensitivity}(c, t) \frac{\partial (1-\operatorname{Specificity}(c,t))}{\partial c} dc \]
The AUC represents the likelihood that diagnostic test outcomes from a randomly chosen pair (one diseased and one non-diseased individual) are arranged correctly.

The cumulative sensitivity (\(C\)) and dynamic specificity (\(D\)) at specific thresholds and time points are defined as:
\[
\operatorname{Sensitivity}_{C}(c, t) = P\left(X_i > c \mid T_i \leq t\right)
\]
\[
\operatorname{Specificity}_{D}(c, t) = P\left(X_i \leq c \mid T_i > t\right)
\]
Here, \(C\) stands for ``Cumulative,'' and \(D\) stands for ``Dynamic.''

The area under the \(C,D\)-curve at time \( t \) is given by:
\[
AUC_{C,D}(t) = P\left(X_i > X_j \mid T_i \leq t, T_j > t, i \neq j\right)
\]

In here we will be using the method that was presented by \cite{Li2018}, and will briefly describe it in here. \cite{Li2018} proposed approach builds upon existing work, particularly the model-based method by \cite{Chambless2006}. \cite{Chambless2006} utilized a semi-parametric Cox model to define \( W_i \) as the conditional probability of \( T_i \leq t \) given \( M_i \) for all subjects.  But this method diverges from this by specifying the weight ($W_{i})$ as the conditional probability of \( T_i \leq t \) given \( M_i \) only for those subjects with \(  \min(T_{i}, C_{i})= Y_i < t \) and \( \delta_i = 0 \); \( W_i \) is either 0 or 1 for others. In here,  $C_{i}$ being a censoring time. Additionally, while \cite{Chambless2006} approach is semi-parametric, this method is fully non-parametric, relying on kernel estimation.

A key advantage of this method lies in its robustness to the tuning parameter, such as bandwidth. Unlike the nearest neighbor method by \cite{Heagerty2000}, which is highly sensitive to the tuning parameter, this method maintains consistent performance across varying conditions.

There are four scenarios to consider when dealing with time-dependent right-censored data:

\begin{enumerate}
    \item if $Y_i > t$, this subject is called a control (or survivor, assuming without loss of generality that the disease condition represents death) at $t$, and the disease status $D_i(t) = 0$. It assigns a weight $W_i = 0$ to this subject. 
    \item if $Y_i \leq t$ and $\delta_i = 1$, this subject called a case (or non-survivor) at time $t$, and the disease status $D_i(t) = 1$. It assigns a weight $W_i = 1$ to this subject.
    \item if $Y_i = t$ and $\delta_i = 0$, then $T_i > Y_i = t$ and the disease status $D_i(t) = 0$. The weight $W_i$ of this subject is $0$. Note that when the time is on a continuous scale, then theoretically the probability of this scenario is zero. 
    \item if $Y_i < t$ and $\delta_i = 0$, the disease status is unknown for this subject, but the probability that this subject is a non-survivor is $P(T_i \leq \tau | Y_i, M_i)$, and the probability that this subject is a survivor is $P(T_i > t | Y_i, M_i)$. For each subject with $Y_i < t$ and $\delta_i = 0$, it defines the weight of this subject to be the probability of being a non-survivor:
\end{enumerate}

\begin{equation}
    W_i = P(T_i \leq t | Y_i, M_i) = 1 - \frac{S_{T}(t | M_i)}{S_{T}(Y_i | M_i)}
\end{equation}

where $S_{T}(t|X) = P(T > t|M)$ denotes the conditional survival distribution of $T$ given the biomarker $M$, which can be estimated using kernel-weighted Kaplan-Meier method with a bandwidth $h$ (\cite{Akritas1994}):

\begin{align}
  \hat{S}_{h}(t | M_{i})=  \hat{P}(T_i > t|M_i) &= \prod_{s \in S, s \leq t} \left(1 - \frac{\sum_j K_h(M_j ,M_i)\mathbb{I}(Y_j = s)\delta_j}{\sum_j K_h(M_j ,M_i)\mathbb{I}(Y_j > s)}\right)
\end{align}

where $S$ is the set of distinct $Y_i$'s with $\delta_i = 1$.  

The sensitivity and specificity can then be estimated in non-iterative, closed expression as:

\begin{align*}
    \widehat{\operatorname{Sensitivity}}(c,t)=\hat{P}(M_i > c|T_i \leq t) &= \frac{\sum_{i=1}^{n} W_i \mathbb{I}(M_i > c)}{\sum_{i=1}^{n} W_i} \\
   \widehat{\operatorname{Specificity}}(c,t)= \hat{P}(M_i \leq c|T_i > t) &= \frac{\sum_{i=1}^{n} (1-W_i) \mathbb{I}(M_i \leq c)}{\sum_{i=1}^{n} (1-W_i)}
\end{align*}

The \textbf{\texttt{tdROC}} package (\cite{Liang_2016_tdROC}) in \textsf{R} (\cite{R}) can be implemented to calculate the time-dependent ROC.

\section{Numerical Studies} \label{sec:numerical_study}

In this simulation, we follow the approach of \cite{Pei_2022}, generating $T_i$ from the following group-specific linear transformation model: 
\[
H(T_i) = \beta_{k,1}X_{i,1} + \beta_{k,2}X_{i,2} + \varepsilon_i, \quad i = 1, 2, \ldots, n; \quad k = 1, 2.
\]
where $H(t) = \log\left(2(e^{4t} - 1)\right)$ and $\varepsilon_i$ follows the extreme-value distribution. In this case, the linear transformation model is equivalent to the Cox proportional hazards model. We generate samples from a two-component Cox proportional hazards model with mixing weights $\pi_1 = \frac{1}{3}, \pi_2 = \frac{2}{3}$, and $\bm{\beta}_1 = (-3, -2)^T$, $\bm{\beta}_2 = (1, 1)^T$. The covariates $X_i$ are generated from a multivariate normal distribution with a mean of zero and a first-order autoregressive structure $\bm{\Sigma} = (\sigma)_{st}$ with $\sigma_{st} = 0.5^{|s-t|}$ for $s, t = 1, 2$. The censoring time is generated from a uniform distribution on $[0, C]$, where $C$ is chosen to achieve censoring proportions of 5\% and 25\%.

We implement our penalized likelihood approach, incorporating smoothly clipped absolute deviation (SCAD) and minimax concave penalty (MCP) alongside the logarithmic scaled (LS) penalties, conducting 100 simulations for each. The initial configuration sets the maximum number of components to $10$, with initial mixing proportions uniformly distributed as $\bm{\pi}^{(0)} = \left(\frac{1}{10}, \frac{1}{10}, \ldots, \frac{1}{10}\right)^\top$. The tuning parameters $\rho, \kappa, \alpha$ are determined as $ c\sqrt{\log(n)/n}$ where $c$ is chosen via the proposed Bayesian Information Criterion (BIC) approach which  is justified in \cite{Pei_2022}.

Table \ref{tab:results1} contrasts the performance of two penalty types, SCAD and MCP penalties, against LS penalty method, when the number of components is correctly identified. The evaluation is carried out over two different sample sizes ($n=600$ and $n=900$), considering three parameters ($\bm{\pi}$, $\bm{\beta}_1$, and $\bm{\beta}_2$) for two components. 

For sample size $n = 600$, the bias and standard deviation values for SCAD and MCP penalties generally show a smaller magnitude of bias compared to the LS penalty, across all parameters ($\pi$, $\beta_1$, and $\beta_2$) for both components. Specifically, for parameter $\pi$, the SCAD penalty exhibits a very small bias and comparable standard deviation to MCP, both significantly outperforming the LS penalty in terms of bias. For $\beta_1$ and $\beta_2$, SCAD and MCP penalties also demonstrate superior bias performance compared to the LS penalty, with SCAD showing a slightly better bias reduction for $\beta_2$ and MCP for $\beta_1$.

For sample size $n = 900$, the differences in performance become more pronounced. The SCAD penalty continues to show reduced bias across most parameters compared to the LS penalty, particularly notable in $\beta_1$ and $\beta_2$ for component 1, where it significantly outperforms the LS penalty. The MCP penalty, while it exhibits a favorable reduction in bias for $\pi$ and $\beta_2$ in component 1, shows a mixed performance in other areas, particularly with a noticeable increase in bias for $\beta_1$ in component 1. However, it generally maintains a lower or comparable level of standard deviation.

Across both sample sizes, the SCAD penalty appears to provide a consistently low bias across all parameters and components, suggesting its effectiveness in parameter estimation under the conditions tested. The MCP penalty, while effective in certain conditions, shows variability in its performance, particularly in the larger sample size. The LS penalty consistently shows higher bias, indicating its relative inefficiency in handling the specific conditions of this study.

These findings suggest that both SCAD and MCP penalties can be more effective than LS penalties in reducing bias in parameter estimation, with SCAD showing particular promise. However, the choice between SCAD and MCP might depend on specific conditions, such as the parameter in question and the sample size. This analysis shows the importance of choosing an appropriate penalty method in statistical modeling, especially in the presence of right-censoring.

\begin{table}[h!]
\centering
\caption{Bias and standard deviation (in parentheses) of parameter estimators for various sample sizes, and penalty types, and $5\%$ right-censoring}
\label{tab:results1}
\begin{tabular}{@{}llccccc@{}}
\toprule
& & &  & Penalty Type \\ \hline 
Sample Size & Component & Parameter & SCAD & MCP & LS\\
\midrule
$n = 600$ &  1 & $\pi$ & $0.00080(0.01527)$ &  $0.00215 (0.01539)$ & $-0.02094(0.01691)$\\
        &   & $\beta_1$ & $0.00113(0.13547)$ & $-0.00760  (0.13285 )$ & $ 0.01753(0.13063)$ \\
        &   & $\beta_2$ & $0.00258(0.11909)$ & $-0.00042  (0.12294)$& $0.01245(0.11759)$ \\ \hline 
        & 2 & $\pi$ & $-0.00080(0.01527)$ & $-0.00215 (0.01539)$& $0.02094(0.01691)$ \\
        &     & $\beta_1$ & $0.00700(0.11787)$ &  $0.00300 (0.11597 )$ & $0.01700(0.11921)$\\
        &      & $\beta_2$ & $0.00967(0.13017)$ & $ 0.00661 (0.13227 )$ & $0.02295(0.128895)$ \\  \hline  \hline 
$n = 900$ &  1 & $\pi$ & $-0.00171(0.03622)$ & $0.00155(0.01281)$ & $-0.01621(0.01355)$ \\
        &    & $\beta_1$ & $0.00873(0.21105)$   & $-0.02163(0.12968)$  &  $0.00856(0.13430)$\\
        &      & $\beta_2$ & $0.00793(0.14603)$ & $-0.01101(0.09043)$ & $0.00477(0.08532)$ \\  \hline 
        &    2 & $\pi$ & $0.00171(0.03622)$ & $-0.00155(0.01281)$ & $0.01621(0.01355)$\\
        &     & $\beta_1$ & $-0.00494(0.08225)$ & $-0.00414(0.08658)$ & $0.00894(0.08983)$ \\
        &     & $\beta_2$ & $0.00540(0.09627)$ &  $0.00120(0.09933)$ & $0.00642(0.09934)$ \\ 
\bottomrule
\end{tabular}
\end{table}

The SCAD penalty emerges as the most reliable across both sample sizes, displaying a strong ability to accurately estimate the number of components under the specified right-censoring conditions. The performance of the MCP penalty, while generally robust, slightly deteriorates with the larger sample size due to a mild increase in overestimation. The LS penalty, although showing a high degree of accuracy at $n=600$, appears to suffer from a loss of precision when the sample size is increased to $n=900$. This suggests that while SCAD and MCP penalties are relatively stable and superior in their estimation capabilities, the LS penalty's effectiveness may be more sensitive to sample size changes, with a tendency towards overfitting that becomes more pronounced as the number of observations grows.

Table \ref{tab:results2} presents the bias and standard deviation of parameter estimators across different sample sizes, penalty types, and with a higher level of right-censoring ($25\%$), we can draw several insights into the performance of SCAD, MCP, and LS penalty methods under these conditions when the number of components is correctly identified. This comparison is made considering two sample sizes ($n=600$ and $n=900$), for three parameters ($\bm{\pi}$, $\bm{\beta}_1$, and $\bm{\beta}_2$) across two components.

For sample size $n = 600$, the results indicate that both SCAD and MCP penalties are effective in minimizing the bias for parameter $\pi$, with extremely low bias values, significantly better than the LS penalty. This suggests that both penalties are robust against right-censoring to a certain extent. For $\beta_1$ and $\beta_2$, the SCAD and MCP methods again show a lower bias compared to the LS penalty, although the differences in bias and standard deviation for these parameters are less pronounced than for $\pi$. Interestingly, for $\beta_2$, both SCAD and MCP exhibit a negative bias, indicating a slight underestimation, which is nonetheless preferable to the positive bias observed with the LS penalty.

Moving to sample size $n = 900$, the performance of the SCAD and MCP penalties in terms of bias reduction becomes even more evident. For parameter $\pi$, both penalties manage to maintain a low bias, markedly lower than that of the LS penalty. This underscores the effectiveness of these penalties in dealing with the challenges posed by increased right-censoring. The parameters $\beta_1$ and $\beta_2$ show a diverse range of bias values, with SCAD and MCP penalties generally exhibiting a negative bias, implying underestimation. However, it's noteworthy that for $\beta_1$, the LS penalty shows a bias closer to zero, suggesting that in some cases, it might offer competitive performance under increased sample sizes.

Across both sample sizes and for all parameters, the standard deviation values are relatively close across the three penalty methods, indicating that the precision of the estimators is somewhat consistent regardless of the penalty applied. However, the bias values suggest that SCAD and MCP penalties provide a more accurate estimation, especially under conditions of higher right-censoring.

In conclusion, the findings suggest that the SCAD and MCP penalties are generally more effective than the LS penalty method in reducing bias in parameter estimation under conditions of $25\%$ right-censoring. This is particularly true for the estimation of $\pi$, where both penalties exhibit superior performance across sample sizes. For $\beta_1$ and $\beta_2$, the benefits of SCAD and MCP penalties are also evident, although the specific advantages vary depending on the parameter and sample size.

\begin{table}[h!]
\centering
\caption{Bias and standard deviation (in parentheses) of parameter estimators for various sample sizes, and penalty types, and $25\%$ right-censoring}
\label{tab:results2}
\begin{tabular}{@{}llccccc@{}}
\toprule
& & &  & Penalty Type \\ \hline 
Sample Size & Component & Parameter & SCAD & MCP & LS\\
\midrule
$n = 600$ &  1 & $\pi$ & $ 0.00003 (0.01604)$ &  $0.00079 (0.01551)$ & $-0.02197 (0.01801)$\\
        &   & $\beta_1$ & $0.00196 (0.15540)$ & $0.00328 (0.15990)$ & $0.0205 (0.15148)$ \\
        &   & $\beta_2$ & $ -0.00237(0.14296)$ & $-0.00314 (0.14572)$& $0.01154(0.13998)$\\ \hline 
        & 2 & $\pi$ & $-0.00003 (0.01604)$ & $-0.00079 (0.01551)$ & $0.02197 (0.01801)$ \\
        &     & $\beta_1$ & $0.01195 (0.12659)$ &  $0.01782 (0.12391)$ & $0.01913 (0.12575)$\\
        &      & $\beta_2$ & $0.02965 (0.12925)$ & $0.00424 (0.12455)$ & $0.02623 (0.13225)$ \\  \hline  \hline 
$n = 900$ &  1 & $\pi$ & $0.00160(0.01175)$ & $0.00069(0.01163)$ & $-0.01718(0.01361)$ \\
        &    & $\beta_1$ & $ -0.02132(0.15314)$ & $ -0.02422(0.14722)$ & $-0.00222(0.14598)$ \\
        &      & $\beta_2$ & $-0.01881(0.10240)$ & $-0.02495(0.10021)$ & $-0.00816(0.10208)$ \\  \hline 
        &    2 & $\pi$ & $-0.00160(0.01175)$ & $-0.00069(0.01163)$ & $ 0.01718(0.01361)$\\
        &     & $\beta_1$ & $0.00754(0.09868)$ & $ 0.00729(0.09977)$ & $0.01195(0.12659)$ \\
        &     & $\beta_2$ & $0.01053(0.10170)$ & $ -0.00167(0.10696)$ & $0.01499(0.10451)$ \\ 
\bottomrule
\end{tabular}
\end{table}

In conclusion, the SCAD penalty exhibits consistently high accuracy across sample sizes, while the MCP penalty shows slight overestimation, which is more pronounced at larger sample sizes. The LS penalty, however, appears to be less reliable, with its accuracy adversely affected as the number of observations increases, but indeed show a high precision for a smaller sample size.

\section{Real survival data experiments} \label{sec:real_data}

We compare the performance of the Cox proportional hazard (PH) model and the mixture Cox PH model with the three type of the penalties, LS, SCAD, and MCP on two datasets from real studies: The Molecular Taxonomy of Breast Cancer International Consortium (METABRIC) and Rotterdam \& German Breast Cancer Study Group (GBSG).

{\bf{Molecular Taxonomy of Breast Cancer International
Consortium (METABRIC)}}

We employ datasets as outlined in (\cite{Katzman_2018_DeepSurv}). METABRIC studies gene and protein levels to find new groups of breast cancer. This helps physicians make better treatment choices. The METABRIC data includes information on genes and health from $1,980$ patients. Out of these, $57.72\%$ passed away from breast cancer, with an average survival time of $116$ months (\cite{Curtis_2012_genomic}). The data was prepared using the Immunohistochemical 4 plus Clinical (IHC4+C) test (\cite{Katzman_2018_DeepSurv}). This test is often used to decide how to treat breast cancer patients (\cite{Lakhanpal_2016_IHC4}). They combined information on four genes (MKI67, EGFR, PGR, and ERBB2) with patients’ health details like hormone treatment, radiotherapy, chemotherapy, ER status, and age at diagnosis. We kept $20\%$ of the data for testing.

{\bf{Rotterdam \& German Breast Cancer Study Group (GBSG)}}

 We utilize datasets as described in \cite{Katzman_2018_DeepSurv}, specifically focusing on the datasets from the Rotterdam Tumor Bank  (\cite{Foekens_2000_Urokinase}) and the German breast cancer study group (GBSG) (\cite{Schumacher_1994_Randomized})
, for our Cox mixture model. It is important to note that our methodology diverges from \cite{Katzman_2018_DeepSurv}; we do not employ its analysis techniques or treatment recommendation system. Instead, our approach leverages these datasets to explore the efficacy of the Cox mixture model in analyzing breast cancer patient data. The Rotterdam dataset comprises information on $1546$ patients with node-positive breast cancer, with observed death times for nearly $90\%$ of these patients. Our validation process utilizes the GBSG dataset, which includes data from $686$ patients in a randomized clinical trial assessing the effects of chemotherapy and hormone treatment on survival rates, with $56\%$ of this data censored

In our study, we evaluated the performance of four predictive methods for both of the datasets: the Cox proportional hazards (Cox PH) model, Cox PH mixture (LS), Cox PH mixture (SCAD), and Cox PH mixture (MCP). We compared these methods based on the area under the ROC curve (AUC) metric, on test dataset, to assess their predictive accuracy and generalization capability.


Table \ref{tab:AUC_data_METABRIC} presents the AUC values on the test set, evaluating the predictive accuracy of four different models across four time points: 18, 20, 22, and 24 years. The table includes the one-class Cox PH model and its mixture model variants with LS, SCAD, and MCP penalties. The estimated number of selected components for each mixture model is represented by $\hat{K}$. In this context, a higher AUC value indicates better model performance in distinguishing between positive and negative instances. 

For the 18-year time point, the Cox PH mixture model with SCAD penalty shows the highest AUC value (0.729), while at the 20-year time point, the MCP penalty model achieves the highest AUC (0.731). At 22 years, the LS penalty model demonstrates superior performance with an AUC of 0.726, and at 24 years, the MCP penalty model again outperforms the others with the highest AUC value (0.744). These results indicate that different penalties may excel at specific time points, but the MCP penalty model generally demonstrates strong and consistent predictive ability.

\begin{table}[h!]
\centering
\caption{AUC values for different models across time points for METABRIC dataset}
\begin{tabular}{@{}lccccc@{}}
\toprule
Model & $\hat{K}$ &  18 years & 20 years & 22 years & 24 years \\ \midrule
Cox PH & - &  0.720	&0.708	&0.637	&0.635 \\
Cox PH mixture(LS) & 2 & 0.706	& 0.718	& {\bf 0.726}	& 0.731 \\
Cox PH mixture(SCAD) & 2 & {\bf 0.729 } &	0.722 &	0.679	 & 0.731\\
Cox PH mixture(MCP) & 2 & 0.717 &	{\bf 0.731}&	0.717&	{\bf 0.744} \\ \bottomrule
\end{tabular}
\label{tab:AUC_data_METABRIC}
\end{table}

\begin{figure}[h!]
\centering
\subfloat[]{%
  \includegraphics[width=0.45\textwidth]{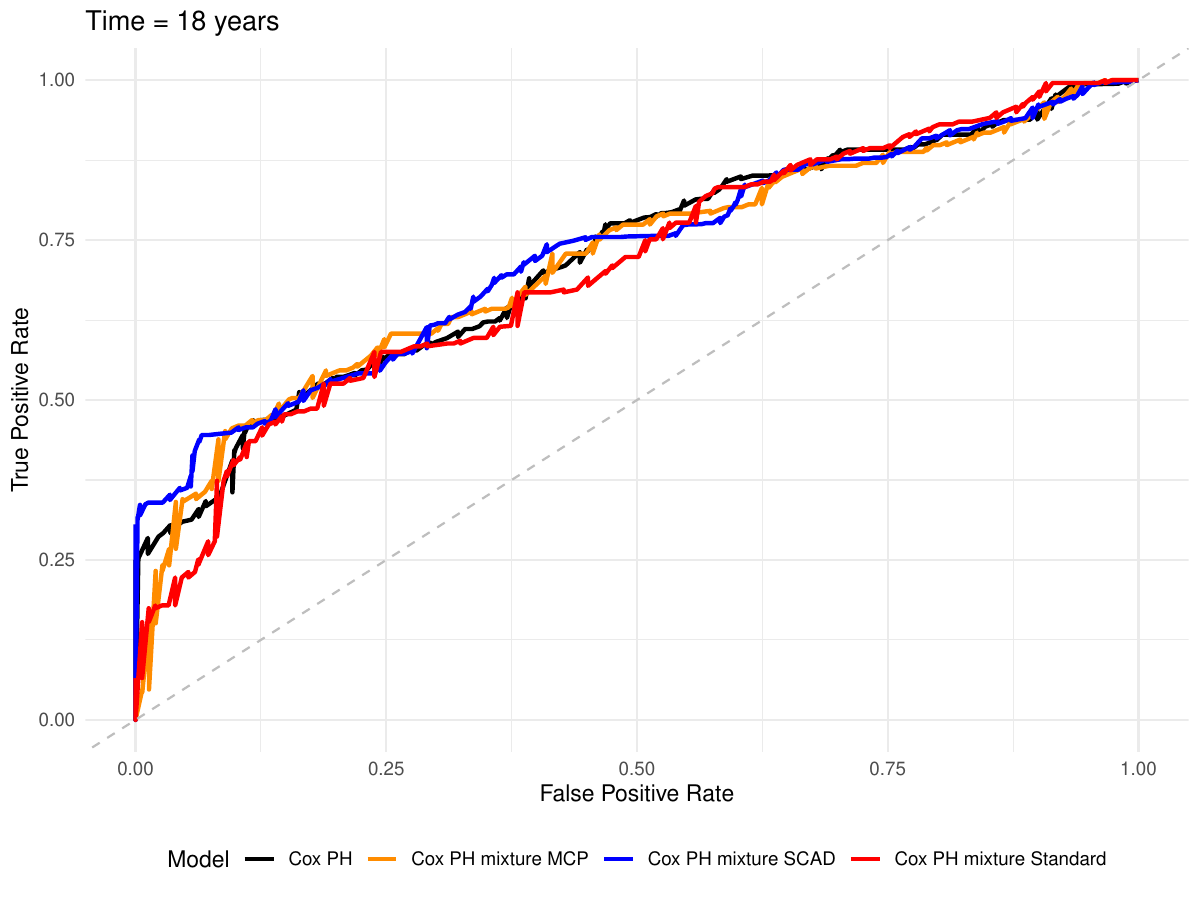}}%
\hfill
\subfloat[]{%
  \includegraphics[width=0.45\textwidth]{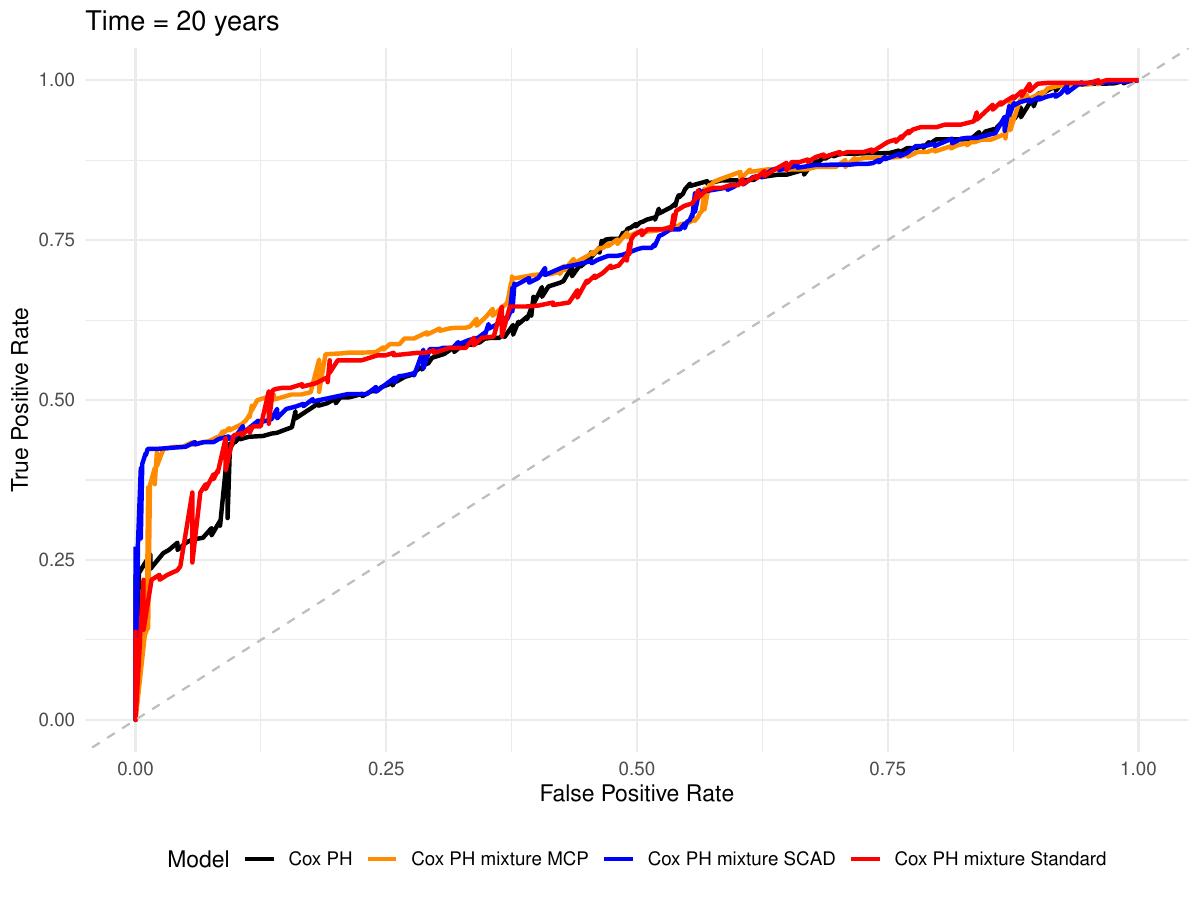}}\\[10pt]
\subfloat[]{%
  \includegraphics[width=0.45\textwidth]{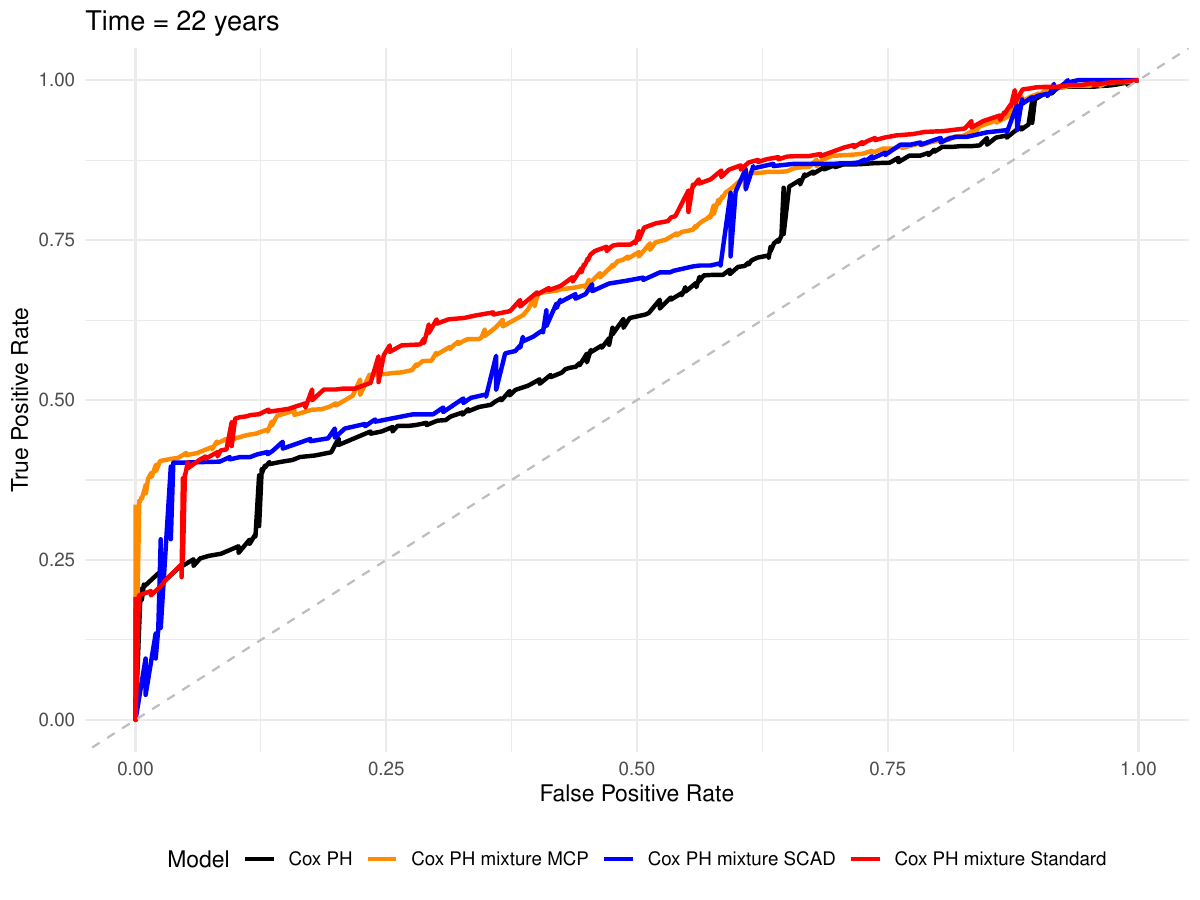}}%
\hfill
\subfloat[]{%
  \includegraphics[width=0.45\textwidth]{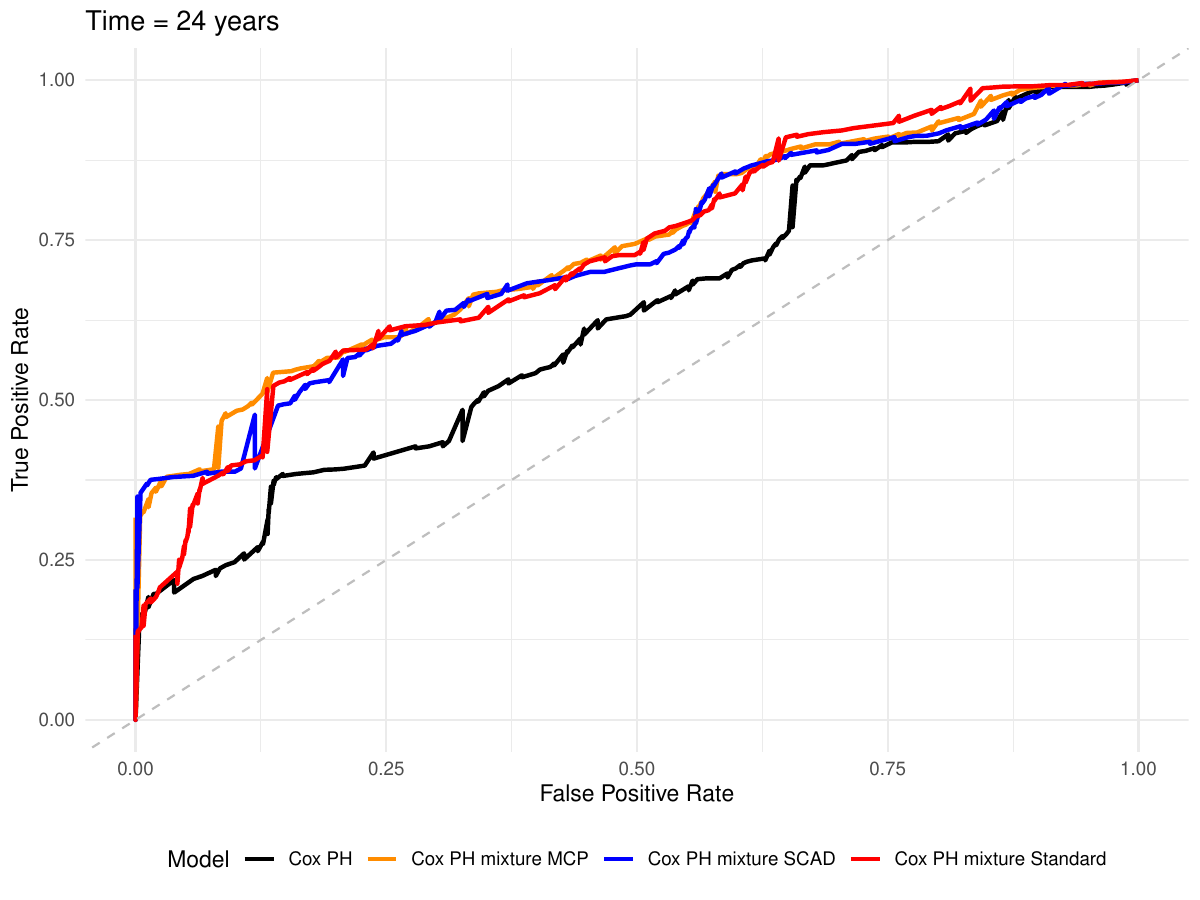}}
\caption{ROC curves of the test dataset for assessing the predictive ability of the one-class Cox PH model and the Cox PH mixture model with LS, SCAD, and MCP penalties over different times for the METABRIC dataset.}
\label{fig:AUC_data_METABRIC}
\end{figure}

Figure \ref{fig:AUC_data_METABRIC} presents the ROC curves for the test dataset at four different time points: 18, 20, 22, and 24 years. These curves assess the predictive ability of the one-class Cox PH model and its mixture model variants with LS, SCAD, and MCP penalties. The curves demonstrate the trade-off between the true positive rate and the false positive rate for the different models. A higher curve towards the top-left corner indicates better model performance.

The results reveal that the mixture models generally yield improved predictions compared to the one-class model, with the SCAD and MCP penalties showing superior performance at specific time points. For example, at 18 years, the Cox PH mixture model with SCAD penalty achieves the highest AUC of \(0.729\), outperforming both the MCP penalty model (\(0.717\)) and the one-class model (\(0.720\)). Similarly, at 24 years, the MCP penalty model achieves the highest AUC of \(0.744\), which is substantially higher than \(0.635\) achieved by the one-class model and comparable to \(0.731\) by the LS penalty model. These results emphasize the enhanced predictive accuracy of the SCAD and MCP penalty models in distinguishing survival likelihoods across different time points.


Table \ref{tab:AUC_data_GBSG} provides a comprehensive comparison of predictive accuracy, as measured by AUC, for four distinct models applied to the GBSG dataset over four specified time intervals: 18, 24, 30, and 36 months. This evaluation encompasses the one-class Cox PH model alongside its three mixture model variations, which are differentiated by their penalty terms. The data demonstrates that the Cox PH mixture model with the SCAD penalty outperforms the other models at every time point, achieving the highest AUC values consistently across all four intervals. Specifically, the table highlights the superior performance of the SCAD-penalized Cox PH mixture model, showing its enhanced predictive accuracy within the given dataset.


Notably, the Cox PH mixture model with LS penalty selected a relatively modest number of components, specifically 2, indicating a cautious approach towards model complexity and overfitting. Despite their restrained model complexity, these models deliver competitive AUC values, illustrating a fine balance between simplicity and accuracy. This balance is a critical aspect of statistical modeling, where the aim is not only to enhance predictive performance but also to maintain model interpretability and generalizability.

Conversely, the Cox PH mixture models incorporating the SCAD and MCP penalties opts for a significantly higher number of components, with 10 and 9 selected components, respectively, suggesting a more complex model structure. This increased complexity could potentially lead to overfitting; however, the SCAD model not only averts this pitfall but also achieves the highest AUC values across all time points. This outcome underscores the SCAD model's efficacy in capturing the underlying patterns within the GBSG dataset more effectively than its counterparts. The SCAD's design to penalize model complexity adaptively plays a pivotal role in its success, allowing it to enhance predictive accuracy without succumbing to overfitting. This insight is particularly relevant in the context of medical research, where the accuracy of prognostic models can directly influence clinical decision-making and patient outcomes.

\begin{table}
\centering
\caption{AUC values for different models across time points for GBSG dataset}
\begin{tabular}{@{}lccccc@{}}
\toprule
Model & $\hat{K}$ &  18 months & 24 months & 30 months & 36 months \\ \midrule
Cox PH & - &  0.684 &	0.682 &	0.701 &	0.708
  \\
Cox PH mixture(LS) & 2 & 0.705	& 0.696 &	0.716	& 0.714 \\
Cox PH mixture(SCAD) & 10 & {\bf 0.721} &	{\bf 0.708}	& {\bf 0.725} &	{\bf 0.725}\\
Cox PH mixture(MCP) & 9 & 0.709 &	0.701 &	0.721 &	0.721 \\ \bottomrule
\end{tabular}
\label{tab:AUC_data_GBSG}
\end{table}

\begin{figure}[h!]
\centering
\subfloat[]{%
  \includegraphics[width=0.45\textwidth]{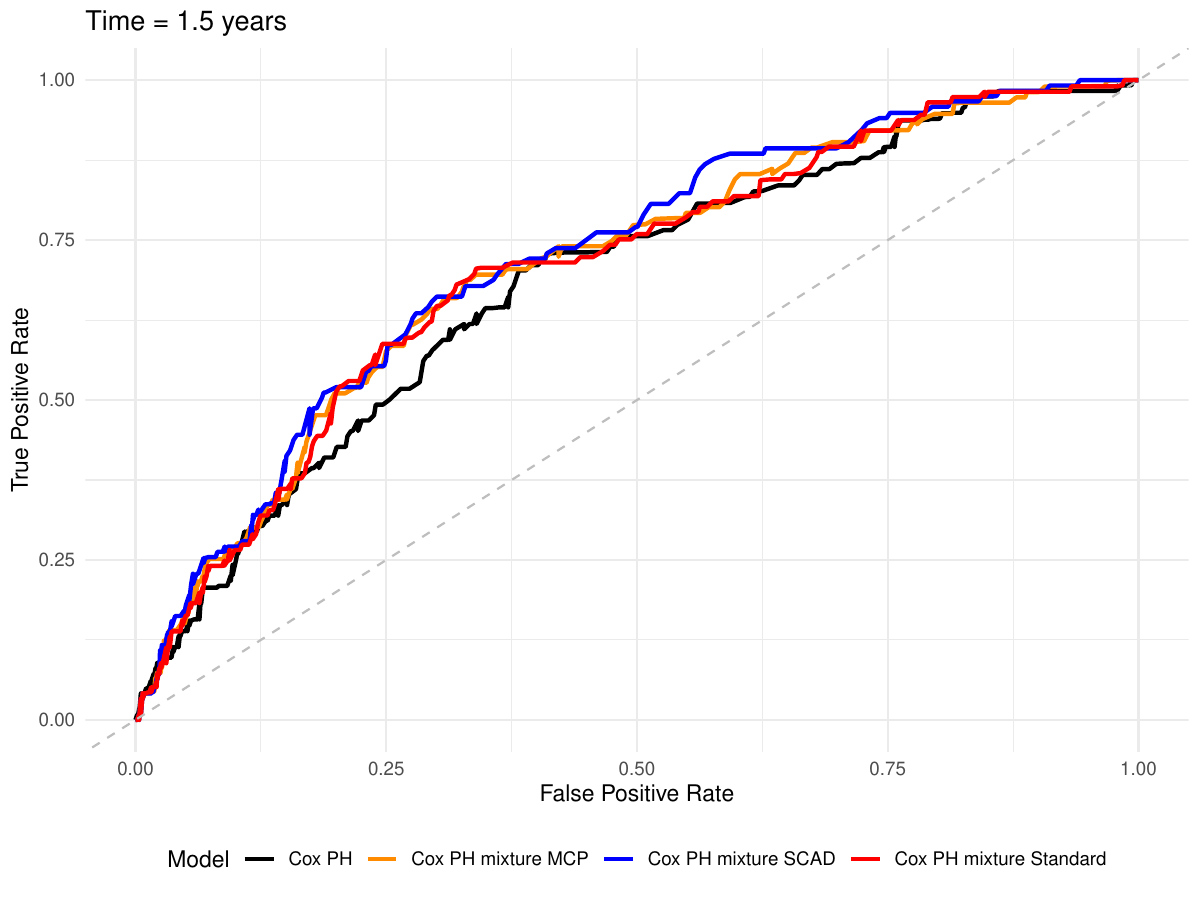}}%
\hfill
\subfloat[]{%
  \includegraphics[width=0.45\textwidth]{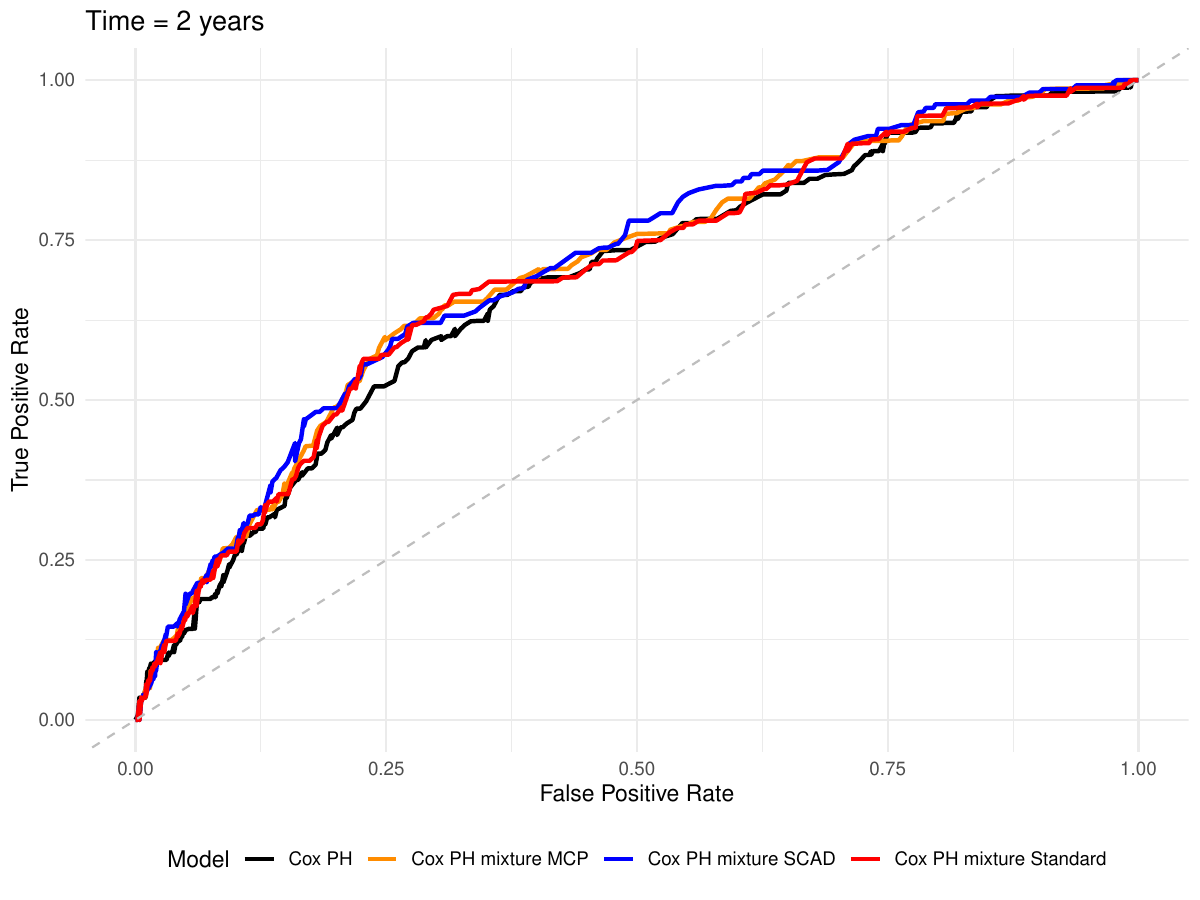}}\\[10pt]
\subfloat[]{%
  \includegraphics[width=0.45\textwidth]{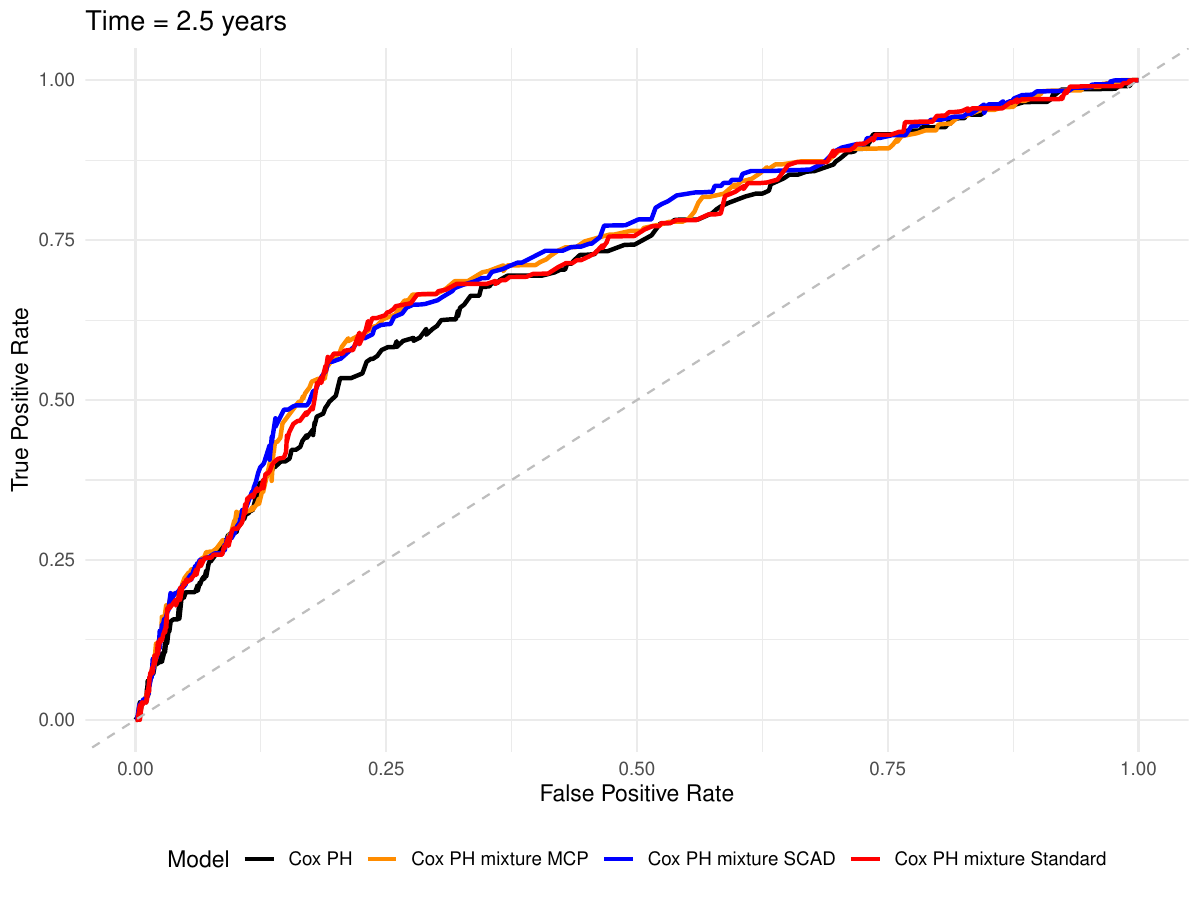}}%
\hfill
\subfloat[]{%
  \includegraphics[width=0.45\textwidth]{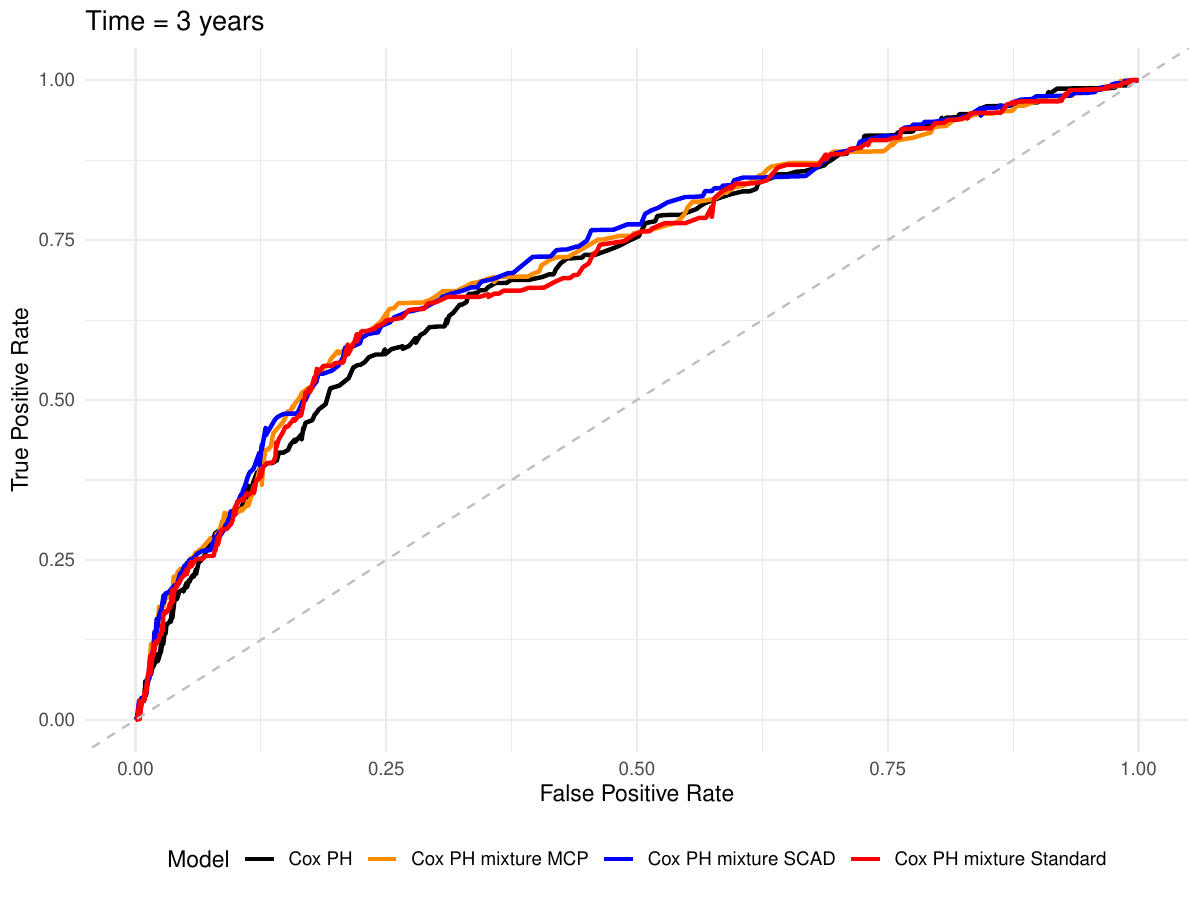}}
\caption{ROC curves of the test dataset for assessing the predictive ability of the one-class Cox PH model and the Cox PH mixture model with LS, SCAD, and MCP penalties over different times for the GBSG dataset.}
\label{fig:AUC_data_GBSG}
\end{figure}

Figure \ref{fig:AUC_data_GBSG} illustrates the ROC curves for the test dataset, examining the predictive accuracy of different Cox PH model variants on GBSG dataset at various follow-up times: 18, 24, 30, and 36 months. Each subfigure, (a) to (d), corresponds to these time points and plots the true positive rate against the false positive rate for the one-class Cox PH model and its mixture models with LS, SCAD, and MCP penalties.

The ROC curves indicate that the SCAD-penalized mixture model consistently achieves the best predictive performance across all time points, as reflected by its highest AUC values. For example, at 18 months, the SCAD-penalized model achieves an AUC of \(0.721\), outperforming the MCP-penalized model (\(0.709\)) and the LS-penalized model (\(0.705\)). Similarly, at 36 months, the SCAD model remains superior with an AUC of \(0.725\), compared to the MCP model (\(0.721\)) and the LS model (\(0.714\)). These results emphasize the effectiveness of the SCAD penalty in enhancing the predictive accuracy of survival likelihoods over multiple follow-up intervals.


The ROC curves are color-coded to distinguish between the models, providing a visual assessment of their predictive performances. A model's predictive power is considered better the closer its curve follows the left-hand border and then the top border of the ROC space. As depicted, the Cox PH mixture model with the MCP penalty consistently outperforms the other models at all time points, which is in agreement with the AUC values presented in Table \ref{tab:AUC_data_GBSG}. The superiority of the MCP model is visually represented by its curves being the uppermost in each panel, denoting the highest true positive rates for given levels of false positive rates and thereby confirming its enhanced predictive accuracy within the GBSG dataset.

\section{Conclusion} \label{sec:conclusion}

This study highlights the limitations of traditional survival analysis methods, especially when dealing with heterogeneous populations. Traditional models, such as the Cox proportional hazards model, often operate under the assumption that a single regression model can properly represent the conditional hazard function across a heterogeneous population. However, this assumption fails to hold in the presence of covariate effect heterogeneity, which naturally leads to the formation of distinct subgroups within the population. This realization highlights the necessity for a more nuanced approach that can account for such heterogeneity. \cite{Pei_2022} addressed this limitation by introducing a version of the Cox proportional hazards model that incorporates heterogeneity effects using a LS penalty term. However, this approach has been critiqued for the penalty term's bias (\cite{Huang_2017}). In contrast, \cite{Fan_2001} highlighted that an ideal penalty function for survival analysis models, should posses three fundamental properties: unbiasedness, sparsity, and continuity. 

To address these challenges, this paper has proposed dynamic penalty functions, such as the SCAD and MCP penalties, into the Cox model to address biasedness. This modification facilitates the handling of heterogeneous time-to-event data and obviates the need for predefined latent class specifications. By enabling a more precise and theoretically grounded determination of the optimal number of mixture components, this approach enhances the accuracy of survival analysis in heterogeneous datasets.

The proposed component selection method, leveraging the strengths of SCAD and MCP penalties, alongside a modified expectation--maximization (EM) algorithm, marks a shift towards more adaptable and accurate survival analysis techniques. Through comprehensive simulation studies and real data analyses, the effectiveness and superiority of these methods have been convincingly demonstrated, offering promising avenues for future research and application.

\section*{Declarations}
\subsection*{Funding}
No funding was received for this research.
\subsection*{Competing Interests}
The authors declare no competing interests.

\subsection*{Ethics Approval}
This study did not involve human participants, animal subjects, or personally identifiable data, and therefore did not require formal ethics approval.


\subsection*{Consent to Participate and Publish}
Not applicable.


\subsection*{Availability of Data}

Both datasets used in this study are publicly available. The METABRIC and GBSG dataset can be accessed at \url{https://github.com/jaredleekatzman/DeepSurv/blob/master/experiments/data}.









\begin{appendices}

\section{Derivation of the SCAD penalty term} \label{appendixA}
Consider 
\begin{equation}
\ell_P (\kappa,\bm{\theta}) = \sum\limits_{i=1}^{n} \sum\limits_{k=1}^{K} s_{ik} \left[ \log(\pi_k) + \log f_k (Y_i, \delta_i|\mathbf{X}_i) \right] -  n \, \kappa \sum\limits_{k=1}^{K} [\log (\varepsilon + p^{\text{SCAD}}_{\kappa,a}(\pi_k)) - \log(\varepsilon)]
\label{eq:eq:appendix_1}
\end{equation}

where $p^{\text{SCAD}}_{\kappa,a}(\pi_k))$ is the SCAD penalty function as proposed by \cite{Fan_2001}. This function is effectively described by its derivative:
\begin{equation}
    p'^{\, \text{SCAD}}_{\kappa,a}(\pi) = I(\pi \leq \kappa) + \frac{(a\kappa - \pi)_+}{(a - 1)\kappa} I(\pi > \kappa);
\label{eq:appendix_2}
\end{equation}

In our approach, we utilize local linear approximation (\cite{zou_2008}) as the primary technique for achieving shrinkage in the mixture probability, which simultaneously ensures stability. That is, 
\begin{equation}
   \log\left( \varepsilon+ p^{\text{SCAD}}_{\kappa,a}(\pi_{k}) \right) \approx   \log\left( \varepsilon+ p^{\text{SCAD}}_{\kappa,a}(\pi^{(0)}_{k}) \right) + \left[\frac{p'^{\, \text{SCAD}}_{\kappa,a}(\pi^{(0)}_{k})}{\varepsilon+p^{\text{SCAD}}_{\kappa,a}(\pi^{(0)}_{k}) }  \right] \left(\pi_{k}-\pi^{(0)}_{k} \right)
   \label{eq:appendix_3}
\end{equation}
and 

\begin{equation}
   \log\left(\pi_{k} \right) \approx   \log\left(\pi^{(0)}_{k}\right) + \frac{1}{\pi^{(0)}_{k}} \left(\pi_{k}- \pi^{(0)}_{k} \right)
      \label{eq:appendix_4}
\end{equation}

Given the current 
  $\bm{\pi}=({\pi}_{1}^{(0)},\ldots,{\pi}_{K}^{(0)})$ for $\bm{\pi}$, then 

  \begin{equation}
      \frac{\partial}{\partial \pi_{k}} \left\{\sum\limits_{i=1}^{n} \sum\limits_{k=1}^{K} s_{ik} \log(\pi_k) -  n \, \kappa \sum\limits_{k=1}^{K} [\log (\varepsilon + p^{\text{SCAD}}_{\kappa,a}(\pi_k))]- \rho_{1} \left( \sum\limits_{k=1}^{K} \pi_{k} -1 \right) \right\} =0 
      \label{eq:appendix_5}
  \end{equation}

  Substituting (\ref{eq:appendix_3}) and (\ref{eq:appendix_4}) into (\ref{eq:appendix_5}), we get,

\begin{equation*}
    \begin{aligned}
        \frac{\partial}{\partial \pi_{k}} &\left\{\sum_{i=1}^{n} \sum_{k=1}^{K} s_{ik} \log(\pi^{(0)}_{k}) + \frac{1}{\pi^{(0)}_{k}} (\pi_{k}- \pi^{(0)}_{k}) -  \right. \\
        & \left. n \kappa \sum_{k=1}^{K} \log(\varepsilon+ p^{\text{SCAD}}_{\kappa,a}(\pi^{(0)}_{k})) + \left[\frac{p'^{\, \text{SCAD}}_{\kappa,a}(\pi^{(0)}_{k})}{\varepsilon+p^{\text{SCAD}}_{\kappa,a}(\pi^{(0)}_{k}) }  \right] (\pi_{k}-\pi^{(0)}_{k})- \rho_{1} \left( \sum_{k=1}^{K} \pi_{k} -1 \right) \right\} = 0 
    \end{aligned}
\end{equation*}
  
Now taking the partial derivative with respect to ${\pi_j}$ with $$ \frac{\partial \pi_k}{\partial \pi_j} = \delta_{jk} = 
\begin{cases} 
0, & \text{if } j \neq k \\
1, & \text{if } j = k 
\end{cases}$$
gives, 

\begin{equation}
      \sum\limits_{i=1}^{n} s_{ij} \frac{1}{\pi^{(0)}_{j}} -  n \, \kappa \left[\frac{p'^{\, \text{SCAD}}_{\kappa,a}(\pi^{(0)}_{j})}{\varepsilon+p^{\text{SCAD}}_{\kappa,a}(\pi^{(0)}_{j}) }  \right] - \rho_{1} =0 
      \label{eq:appendix_6}
  \end{equation}

 Multiplying in by  $\pi^{(0)}_{j}$ gives, 
 \begin{equation}
       \sum\limits_{i=1}^{n} s_{ij}-  n \, \kappa \left[\frac{p'^{\, \text{SCAD}}_{\kappa,a}(\pi^{(0)}_{j}) \, \pi^{(0)}_{j}}{\varepsilon+p^{\text{SCAD}}_{\kappa,a}(\pi^{(0)}_{j}) }  \right] - \rho_{1} \, \pi^{(0)}_{j}=0 
\label{eq:appendix_7}
 \end{equation}
 Now, summing over $K$, gives
 \begin{equation}
       \sum\limits_{j=1}^{K} \sum\limits_{i=1}^{n} s_{ij}-  n \, \kappa \sum\limits_{j=1}^{K} \left[\frac{p'^{\, \text{SCAD}}_{\kappa,a}(\pi^{(0)}_{j}) \, \pi^{(0)}_{j}}{\varepsilon+p^{\text{SCAD}}_{\kappa,a}(\pi^{(0)}_{j}) }  \right] - \rho_{1} \, \sum\limits_{j=1}^{K} \pi^{(0)}_{j}=0 
 \end{equation}

 Since $ \sum\limits_{j=1}^{K} \sum\limits_{i=1}^{n} s_{ij}=n$ and $\sum_{j=1}^{K} \pi^{(0)}_{j}=1$, then 
  \begin{equation}
       n-  n \, \kappa \sum\limits_{j=1}^{K} \left[\frac{p'^{\, \text{SCAD}}_{\kappa,a}(\pi^{(0)}_{j}) \, \pi^{(0)}_{j}}{\varepsilon+p^{\text{SCAD}}_{\kappa,a}(\pi^{(0)}_{j}) }  \right] - \rho_{1} =0 
 \end{equation}
 Therefore, 
 \begin{equation}
          \rho_{1}=  n-  n \, \kappa \sum\limits_{j=1}^{K} \left[\frac{p'^{\, \text{SCAD}}_{\kappa,a}(\pi^{(0)}_{j}) \, \pi^{(0)}_{j}}{\varepsilon+p^{\text{SCAD}}_{\kappa,a}(\pi^{(0)}_{j}) }  \right] 
 \end{equation}

 Substituting $\rho_{1}$ in (\ref{eq:appendix_6}), gives 
 \begin{equation}
 \pi_{k}^{(1)}= \frac{\sum\limits_{i=1}^{n} s_{ki}}{ n-n \kappa \sum\limits_{k=1}^{K} \frac{p'^{\, \text{SCAD}}_{\kappa,a}(\pi^{(0)}_{k}) \, \pi^{(0)}_{k}}{\varepsilon+p^{\text{SCAD}}_{\kappa,a}(\pi^{(0)}_{k}) }  + n \kappa  \frac{p'^{\, \text{SCAD}}_{\kappa,a}(\pi^{(0)}_{k})}{\varepsilon+p^{\text{SCAD}}_{\kappa,a}(\pi^{(0)}_{k}) }}
 \end{equation}




\end{appendices}


\bibliography{sn-bibliography}

\end{document}